\documentclass{INTERSPEECH2023}
\usepackage{amsmath,graphicx,float}
\usepackage{bm, amssymb, booktabs, multirow, url, cite, color, amsmath}

\interspeechcameraready

\title{Phonetic-assisted Multi-Target Units Modeling for Improving Conformer-Transducer ASR system}

\name{Li Li$^{1}$, Dongxing Xu$^{2}$, Haoran Wei$^{3}$, Yanhua Long$^{1}$\sthanks{\ Yanhua Long is the corresponding author, she is also with the Key Innovation Group of Digital Humanities Resource and Research, Shanghai Normal University. The work is supported by the National Natural Science Foundation of China (Grant No.62071302).}}

\address{
  $^1$Shanghai Engineering Research Center of Intelligent Education and Bigdata, \\
     Shanghai Normal University, Shanghai, China\\
  $^2$Unisound AI Technology Co., Ltd., Beijing, China\\
  $^3$Department of ECE, University of Texas at Dallas, Richardson, TX 75080, USA}
\email{lili\_a0@163.com, xudongxing@unisound.com, haoran.wei@utdallas.edu, yanhua@shnu.edu.cn}

\begin{document}
\ninept
\maketitle

\begin{abstract}

Exploiting effective target modeling units is very important and has always been
a concern in end-to-end automatic speech recognition (ASR). In this work, we propose a phonetic-assisted
multi-target units (PMU) modeling approach, to enhance the
Conformer-Transducer ASR system in a progressive representation learning manner.
Specifically, PMU first uses the pronunciation-assisted
subword modeling (PASM) and byte pair encoding (BPE) to produce
phonetic-induced and text-induced target units separately; Then, three new
frameworks are investigated to enhance the acoustic encoder, including a
basic PMU, a paraCTC and a paCTC, they
integrate the PASM and BPE units at different levels for CTC and transducer
multi-task training.  Experiments on both LibriSpeech and
accented ASR tasks show that, the proposed PMU significantly outperforms
the conventional BPE, it reduces the WER of
LibriSpeech clean, other, and six accented ASR testsets
by relative 12.7\%, 4.3\% and 7.7\%, respectively.

\end{abstract}

\section{Introduction}
\label{sec:intro}

The Conformer-Transducer (ConformerT) has achieved state-of-the-art results
in many ASR tasks \cite{conformer, ct, li2020comparison, loss} because of its
perfect inheritance of the advantages of conformer and transducer.
It captures both local and global features by combining the convolution module
and transformer in a parameter-efficient way. Together with the natural streaming
property of transducer, ConformerT has become increasingly appealing in
recent end-to-end (E2E) ASR systems.

As many E2E ASR systems, exploring effective target modeling units for ConformerT
is also very crucial. The main types of E2E ASR
target units can be divided into the text-induced units and phonetic-induced ones.
The character, word and subword are all text-induced units and have been
extensively studied \cite{comparable, comparison, state, advancing, subwordre, rao2017exploring}. Compared
with character, subword can avoid too long output sequence and dependency,
which reduces the difficulty of modeling and decoding \cite{state}. Many
subword modeling techniques have been proposed: the byte pair encoding (BPE) \cite{bpe},
WordPieceModel (WPM) \cite{wpm} and unigram language model (ULM) \cite{unigram}, etc.
However, all of these techniques are purely text-induced without
any access to the underlying phonetic/pronunciation information which is the
key of ASR. The syllable, phoneme \cite{comparable, comparison, phoneme, phone} belong
to the phonetic-induced target units, they enable the model to learn
better phonetic patterns of a language, however, an additional pronunciation lexicon
is required during both model training and inference. Therefore, how to
well exploit the information in both text-induced and phonetic-induced target units
become very important and fundamental.

In the literature, several recent works have been proposed to combine
the text and phonetic information for building better E2E ASR system.
Such as, \cite{investigation} proposed a hybrid target unit of
syllable-char-subword in a joint CTC/Attention multi-task learning
for the Mandarin ASR system; While in \cite{pasm}, a
pronunciation assisted subword modeling (PASM) was introduced to
extract ASR target units by exploring their acoustic structure
from the pronunciation lexicon. In addition, \cite{hierarchical} tried
to exploit the text and underlying phonetic
information in acoustics in another way,
the authors used a set of hierarchically increasing text units to the CTC modeling
of intermediate Transformer encoder layers. All these works have been
verified to be effective for improving current E2E ASR systems.

Motivated by the PASM and work in \cite{hierarchical}, this study aims to
improve the Conformer-Transducer ASR system  by proposing a
phonetic-assisted multi-target units modeling (PMU) approach. The PMU is designed
to learn information from both the phonetic-induced PASM and text-induced
BPE units, using three new acoustic modeling frameworks as follows:
1) Basic PMU. The ConformerT is trained with both CTC and transducer
losses, but assigning PASM and BPE units to them respectively;
2) paraCTC. With the same BPE units as in 1) for transducer,
a parallel CTC loss with both PASM and BPE as Conformer encoder's
target units is taken as the auxiliary task of ConformerT model training;
3) paCTC. Different from 1) and 2), paCTC adopts a phonetic conditioned
acoustic encoder, by using the PASM and BPE units in a interactive manner
to the CTC loss of different intermediate Conformer encoders.
From the experiments that conducted on LibriSpeech and CommonVoice datasets,
we see that the standard ConformerT is significantly improved by our proposed PMU,
up to relative 4.3\% to 12.7\% WER reductions are achieved on the LibriSpeech clean and other,
or the six accented English test sets of CommonVoice.

\begin{figure*}[t]
\centering
\setlength{\abovecaptionskip}{0.2cm}
\includegraphics[width=5.6in,height=3in]{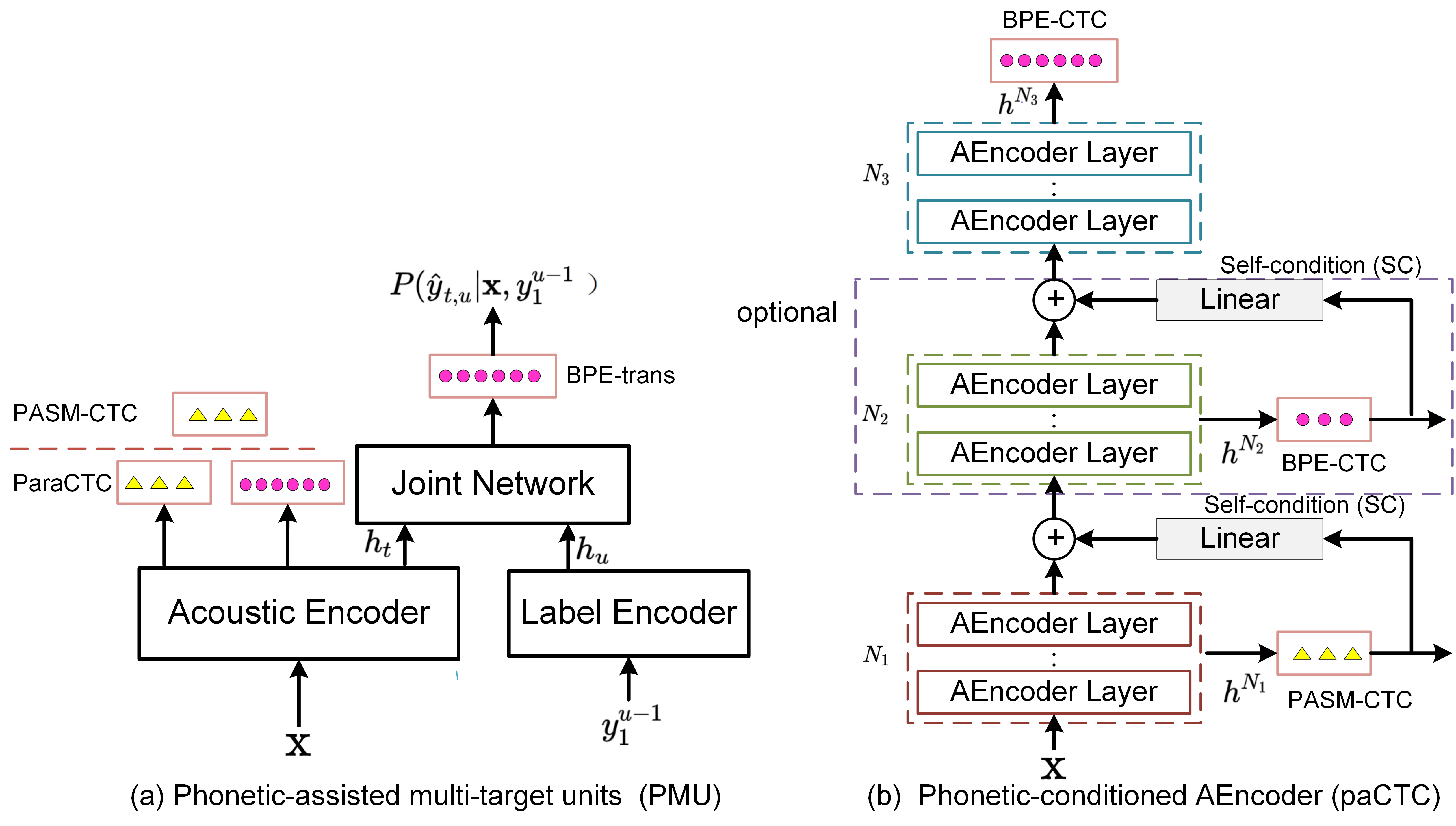}
\caption{Structure of (a) the proposed phonetic-assisted multi-target units (PMU) and (b) its phonetic-conditioned AEncoder (paCTC).}
\label{fig:res}
\vspace{-0.3cm}
\end{figure*}

\section{Conformer-Transducer}
\label{sec:transducer}

The Conformer-Transducer (ConformerT) was first proposed in \cite{conformer, ct}.
It can be trained using the end-to-end RNN-T loss \cite{rnntloss} with
a label encoder and a Conformer-based acoustic encoder (AEncoder).
The architecture is illustrated in Fig.\ref{fig:res} (a).
Given an input acoustic feature with $T$ frames as $\mathbf{x}$  = ($\mathbf{x} _1$,\dots,$\mathbf{x} _T$),
and its transcription label sequence of length $U$ as $\mathbf{y}$  = ($y_1,\dots,y_U$).
The AEncoder first transforms $\mathbf{x}$ into a high representation $h_{t}, t\le T$,
and the label encoder, acting as a language model, produces a representation
$h_{u}$ given its previous emitted label sequence $y_{1}^{u-1}$.
Then, $h_t$ and $h_u$ are combined using the joint network composed of feed-forward layers and a non-linear function to compute output logits.
Finally, by applying a Softmax to the output logits, we can produce the
distribution of current target probabilities as:
\begin{equation}
\label{eq:eq1}
  P(\hat{y}_{t,u}|\textbf{x},y_{1}^{u-1}) = \texttt{Softmax}(\texttt{Joint}(h_t, h_u ))
\end{equation}

The label $\hat{y}_{t,u}$ can optionally be blank symbol. Removing all the
blank symbols in $\hat{y}_{t,u}$ sequence yields  $\mathbf{y}$.
Given $\mathcal{A}$, the set of all possible alignments $\hat{\mathbf{y} }$
(with blank symbols $\phi$) between input $\mathbf{x}$ and output $\mathbf{y}$,
ConformerT loss function can be computed as the following
negative log posterior:
\begin{equation}
\label{eq:eq2}
  \mathcal{L}_{\texttt{trans}}=-\log{P(\mathbf{y}|\mathbf{x})} = -\log {\sum_{\hat{\mathbf{y} }\in \mathcal{A}}P(\hat{\mathbf{y} }|\mathbf{x})}
\end{equation}

Besides the transducer loss $\mathcal{L}_{\texttt{trans}}$, as in \cite{multitask},
we also jointly train ConformerT with an auxiliary CTC loss $\mathcal{L}_{\texttt{CTC}}$ \cite{ctc} to
learn frame-level acoustic representations and provide supervision to the AEncoder.
The overall ConformerT objective function is defined as:
\begin{equation}
\label{eq:eq3}
  \mathcal{L}_{\texttt{obj}} = \lambda _{\texttt{trans}}\mathcal{L}_{\texttt{trans}}+\lambda _{\texttt{ctc}}\mathcal{L}_{\texttt{CTC}}
\end{equation}
where $\lambda _{\texttt{trans}}, \lambda _{\texttt{ctc}} \in [0,1]$ are
tunable loss weights.

\section{Proposed Methods}
\label{sec:Methods}

Although the PASM \cite{pasm} has been proposed to enhance the extraction of E2E ASR target units,
by leveraging the phonetic structure of acoustics in speech using a pronunciation lexicon,
most current ConformerT ASR systems still only use the purely text-induced subwords, such as BPEs, wordpieces
as their target modeling units \cite{ct, li2020comparison}. This may because constrained by the phonetic
pattern in lexicon, PASM tends to produce short subwords and avoids modeling larger or
full-words with single tokens, the resulting relative small vocabulary size
greatly limits the performance upper bound of PASM. Therefore, in this study, we propose
a phonetic-assisted multi-target units (PMU) modeling to integrate the advantages of both
PASM and BPE units for improving ConformerT in a CTC/transducer multi-task training framework.

The whole architecture of ConformerT with PMU is demonstrated in
Fig.\ref{fig:res}(a), where we use different types of target units for the
CTC and transducer branch. The \texttt{BPE-trans} means using text-induced
BPE units to align the transducer outputs during ConformerT training,
while for the shared acoustic encoder (AEncoder) with CTC branch,
we investigate three new target units modeling methods, as illustrated in the
left part of Fig.\ref{fig:res}(a) and Fig.\ref{fig:res}(b), the first one is the basic
PMU with \texttt{PASM-CTC}, where
only PASM units are taken as the CTC targets, the other two replace \texttt{PASM-CTC}
with a \texttt{paraCTC} and a \texttt{paCTC} separately. All the \texttt{PASM-CTC},
\texttt{BPE-CTC} and \texttt{BPE-trans} are composed of a single fully-connected feed-forward
layer with different target units followed by Softmax function. Given
both PASM and BPE units, the overall objective loss of basic PMU  is defined as,
\begin{equation}
\label{eq:basic}
  \mathcal{L}_{PMU} = \lambda _{\texttt{trans}}\mathcal{L}_{\texttt{BPE-trans}}+\lambda _{\texttt{ctc}}\mathcal{L}_{\texttt{PASM-CTC}}
\end{equation}
Where $\mathcal{L}_{\texttt{BPE-trans}}$ and $\mathcal{L}_{\texttt{PASM-CTC}}$ represent
the loss of transducer and CTC using BPE and PASM units as their targets, respectively.
If we use \texttt{paraCTC} or \texttt{paCTC}, the loss of $\mathcal{L}_{\texttt{PASM-CTC}}$
in Eq.(\ref{eq:basic}) will be replaced by their corresponding CTC loss $\mathcal{L}_{\texttt{paraCTC}}$ and $\mathcal{L}_{\texttt{paCTC}}$, respectively. The details of
how to produce PASM units with a given pronunciation lexicon and training
texts can be found in \cite{pasm}.

\subsection{ParaCTC}
\label{ssec:ParaCTC}

Training a model with CTC loss applied in parallel to the final layer
has recently achieved success \cite{hierarchical1, paractc, inductive, char+}. In our \texttt{paraCTC},
as shown in Fig.\ref{fig:res}(a), we use two different linear layers to transform
the AEncoder representation to BPE and PASM units with $\mathcal{L}_{\texttt{CTC}}(\mathbf{y}_{\texttt{BPE}}|\mathbf{x})$
and $\mathcal{L}_{\texttt{CTC}}(\mathbf{y}_{\texttt{PASM}}|\mathbf{x})$ loss, respectively.
The overall loss function of \texttt{paraCTC} is defined as,
\begin{equation}
\vspace{-0.12cm}
\label{eq:paractc}
  \mathcal{L}_{\texttt{paraCTC}} = \alpha \mathcal{L}_{\texttt{CTC}}(\mathbf{y}_{\texttt{PASM}}|\mathbf{x})+(1-\alpha) \mathcal{L}_{\texttt{CTC}}(\mathbf{y}_{\texttt{BPE}}|\mathbf{x})
\end{equation}
where $\alpha\in(0,1)$,  $\mathbf{y}_{\texttt{PASM}}$ and $\mathbf{y}_{\texttt{BPE}}$ represent
the target units of CTC is PASM and BPE respectively. With Eq.(\ref{eq:paractc}),
the underlying phonetic and text structure information in PASM and BPE are
effectively exploited and combined to boost the AEncoder.

\subsection{PaCTC}
\label{ssec:paCTC}

Different from basic PMU and \texttt{paraCTC}, our proposed \texttt{paCTC}
enhances the AEncoder in a phonetic-conditioned manner, by using the CTC alignments between
$\mathbf{y}_{\texttt{PASM}}$ or $\mathbf{y}_{\texttt{BPE}}$ and the output of
intermediate AEncoder layers. The overview structure of \texttt{paCTC} is shown
in Fig.\ref{fig:res}(b). We first cut the whole AEncoder into
the lower $N_{1}$ and top $N_{3}$ layers. Then, the \texttt{PASM-CTC}
and \texttt{BPE-CTC} joint training are applied to these two AEncoder blocks for aligning
their frame-level outputs $h^{N_{1}}$ and $h^{N_{3}}$ respectively,
using their corresponding loss of $\mathcal{L}_{\texttt{PASM-CTC}}^{N_{1}}$ and
$\mathcal{L}_{\texttt{BPE-CTC}}^{N_{3}}$ as,
\begin{equation}
\vspace{-0.17cm}
\label{eq:pa1}
  \mathcal{L}_{\texttt{paCTC}} = \beta \mathcal{L}_{\texttt{PASM-CTC}}^{N_{1}}+ (1-\beta) \mathcal{L}_{\texttt{BPE-CTC}}^{N_{3}}
\end{equation}
Where $\mathcal{L}_{\texttt{paCTC}}$ is the total \texttt{paCTC} loss and
$\beta \in(0,1)$ is a weight parameter.

Moreover, as illustrated in Fig.\ref{fig:res}(b),
a self-condition(SC) mechanism \cite{sc} is applied to further improve the
AEncoder, by making the subsequent AEncoder layers
conditioned on both the previous layer representation and
the intermediate CTC predictions. The \texttt{Linear} in SC means using a fully-connected
layer to linearly transform the dimension of intermediate CTC predictions to
the same dimension of AEncoder layers. We expect \texttt{paCTC} can
outperform the other two PMU variants, because it integrates both PASM and
BPE advantages in a more effective way, by applying \texttt{PASM-CTC}
on lower AEncoder enables ConformerT to learn better acoustic
representations from the phonetic-induced PASM modeling, while applying
\texttt{BPE-CTC} on the top AEncoder helps to produce more robust
linguistic embeddings.

In addition, inspired by the idea of hierarchically increasing subword units
in \cite{hierarchical}, we also design an optional structure (dashed block in Fig.\ref{fig:res})
in \texttt{paCTC}, by inserting an intermediate \texttt{BPE-CTC} alignment at the middle
AEncoder block with $N_{2}$ layers.
With this optional structure, Eq.(\ref{eq:pa1}) is then modified as
follows:
\begin{equation}
\vspace{-0.15cm}
\label{eq:pa2}
  \mathcal{L}_{\texttt{paCTC}} = \frac{\beta}{2}\left(\mathcal{L}_{\texttt{PASM-CTC}}^{N_{1}}+
  \mathcal{L}_{\texttt{BPE-CTC}}^{N_{2}}\right) +
  (1-\beta) \mathcal{L}_{\texttt{BPE-CTC}}^{N_{3}}
\end{equation}
where $\mathcal{L}_{\texttt{BPE-CTC}}^{N_{2}}$ is the \texttt{BPE-CTC} loss of
intermediate middle AEncoder block.
It's worth noting that the intermediate \texttt{BPE-CTC} and \texttt{PASM-CTC} have the same vocabulary size that is much smaller than the one of
\texttt{BPE-CTC} applied to the top AEncoder block, such as 194 versus 3000.
This \texttt{paCTC} with optional structure can not only leverage
low-level phonetic information to produce better high-level linguistic
targets, but also achieve a progressive representation learning
process which can integrate different types of subwords in a fine-to-coarse
manner. What's more, we explore two different variants of \texttt{paCTC} with optional structure, namely \texttt{paCTC-s} and \texttt{paCTC-us}. \texttt{paCTC-s} means we not only share two SC linear layers, but also share the linear layer parameters of both \texttt{PASM-CTC} and intermediate \texttt{BPE-CTC}, while \texttt{paCTC-us} means not.

\vspace{-0.12cm}
\section{Experiments and Results}
\label{sec:results}

\subsection{Datasets}
\label{ssec:datasets}

Our experiments are performed on two open-source English ASR tasks, one is the
LibriSpeech dataset \cite{librispeech} with 100hrs training data and
its clean and other test sets, the other is an accented
ASR task with data selected from CommonVoice corpus \cite{common}.
Our accented English training data has 150 hours (hrs) of speech,
including Indian, US and England accents and each with 50 hrs.
We construct six test sets to evaluate the proposed methods for accented ASR, including
three in-domain tests with 2 hrs US, 1.92 hrs England and
3.87 hrs Indian accent speech, three out-of-domain test sets with 2 hrs
Singapore, 2.2 hrs Canada and 2 hrs Australia accent speech.

\vspace{-0.2cm}
\subsection{Experimental Setup}
\label{ssec:exps}

All our experiments are implemented using library from
the end-to-end speech recognition toolkit ESPnet \cite{espnet}. We use global mean-variance normalized 80-dimensional log-mel filterbank
as input acoustic features. No data augmentation techniques
and no extra language model are applied.

For the acoustic encoder of ConformerT, we sub-sample the input
features by a factor of 4 using two 2D-convolutional layers,
followed by 12 conformer encoder layers with 2048 feed-forward
dimension and 512 attention dimension with 8 self-attention
heads. For the label encoder, we only use a 512-dimensional LSTM.
The joint network is a 640-dimensional feed-forward network with
tanh activation function. The warmup is set to 25000, and
both label smoothing \cite{labelsmooth} weight and dropout is set to 0.1
for model regularization. The BPE units are generated by
SentencePiece \cite{sentencepiece}, and {fast\_align} \cite{Dyer2013ASF}
is used to produce PASM units with the CMU pronunciation
lexicon$\footnote{http://www.speech.cs.cmu.edu/cgi-bin/cmudict}$.
In Table \ref{tab:libr} and Table \ref{tab:accent}, $\beta$  is set to 0.5 and 0.7, respectively,
$\alpha=0.7$ for the paraCTC, $\lambda _{\texttt{trans}}=\lambda _{\texttt{ctc}}=0.5$
for all the systems with paCTC. All the system performances are evaluated
using word error rates (WER (\%)).

\subsection{Results}
\label{ssec:results}

\subsubsection{Results on Librispeech}
\label{sssec:results}

\begin{table}[!htbp]
\vspace{-1.5em}
\renewcommand\arraystretch{1.15}
\caption{WER(\%) on the clean and other test sets of Libri-100hrs ASR task.
$\textrm{TU}_{ctc}$ and $\textrm{TU}_{trans}$ represent the type of
target units for CTC and transducer in ConformerT, respectively. In paCTC,
system 9-10 use the optional structure with $N_{1}=N_{2}=N_{3}=4$,
while system 8 does not ($N_{2}=0, N_{1}=N_{3}=6$).}
\vspace{-1.em}
  \label{tab:libr}
  \centering
  \scalebox{0.9}{
	\begin{tabular}{c|c|l|l|c|c}
		\toprule
		\multirow{2}{*}{ID} & \multirow{2}{*}{Methods} & \multirow{2}{*}{$\textrm{TU}_{ctc}$}  & \multirow{2}{*}{$\textrm{TU}_{trans}$}   & \multicolumn{2}{c}{Evaluation} \\
		\cline{5-6}
		& & & & Clean & Other\\
		 \midrule
		1 & \multirow{3}{*}{ConformerT} & \multicolumn{2}{l|}{BPE-194} & 11.2 & 30.6  \\
        2 &    & \multicolumn{2}{l|}{BPE-3000} & 11.0 & 29.9  \\
        3 &    & \multicolumn{2}{l|}{PASM-194} & 10.5 & 30.5  \\
         \midrule
        4 & \multirow{7}{*}{PMU}  & PASM-194 & BPE-194 & 10.2 & 30.0 \\
        5 &    & BPE-194 & PASM-194 & 10.7 & 30.2 \\
        6 &    & PASM-194 & BPE-3000 & 10.1 & 28.4 \\
        \cline{3-6}
        7 &    & paraCTC & BPE-3000 & 9.8 & 28.4 \\
        \cline{3-6}
        8 &    & paCTC  & BPE-3000 & 9.7 & 28.4 \\
        9 &    & paCTC-s & BPE-3000 & 9.7 & \textbf{28.3} \\
        10 &    & paCTC-us  & BPE-3000 & \textbf{9.6} & 28.6 \\
		 \bottomrule
	\end{tabular}}
    \vspace{-0.3cm}
\end{table}

We first examine our proposed methods on the clean and other test
sets of Librispeech ASR task. Results are shown in Table \ref{tab:libr}.
System 1 to 3 are our ConformerT baselines, each with its both CTC and
transducer branches using a single type of target units. `BPE/PASM-*' means
using BPE or PASM units with different vocabulary size. In our extensive experiments,
we find 194 and 3000 are the best setups for PASM and BPE on the Libri-100hrs
dataset, respectively. `BPE-194' is used to make a fair comparison with `PASM-194'.
System 4-10 are the ConformerT models trained using our proposed PMU framework with
three different variants: the basic PMU (system 4-6), PMU with \texttt{paraCTC} (system 7)
and PMU with different structure of \texttt{paCTC} (system 8-10).

\begin{table*}[!ht]
\vspace{-0.3cm}
\renewcommand\arraystretch{0.9}
\caption{WER(\%) on the in-domain and out-of-domain test sets on accented CommonVoice ASR task.
 In the paCTC, setup 9-10 use the optional structure, while setup 7-8 do not.}
  \label{tab:accent}
  \centering
  \scalebox{0.95}{
	\begin{tabular}{c|c|l|l|c|c|c|c|c|c|c}
		\toprule
		\multirow{2}{*}{ID} & \multirow{2}{*}{Methods} & \multirow{2}{*}{$\textrm{TU}_{ctc}$}  & \multirow{2}{*}{$\textrm{TU}_{trans}$}   & \multicolumn{3}{c|}{In-domain}  & \multicolumn{3}{c|}{Out-of-domain} & \multirow{2}{*}{Overall} \\
		\cline{5-10}
		& & & & England & Indian & US & Australia & Canada & Singapore & \\
		 \midrule
		1 & \multirow{3}{*}{ConformerT} & \multicolumn{2}{l|}{BPE-3000} & 21.9 & 26.8 & 18.7 & 24.1 & 17.1 & 35.4 & 24.4  \\
        2 &    & \multicolumn{2}{l|}{BPE-205} & 21.7 & 25.0 & 17.6 & 24.5 & 16.2 & 33.8 & 23.4  \\
        3 &    & \multicolumn{2}{l|}{PASM-205}& 21.6 & 24.7 & 17.8 & 24.6 & 16.6 & 35.1 & 23.9  \\
        \midrule
        4 & \multirow{7}{*}{PMU}  & PASM-205 & BPE-205 & 21.4 & 24.7 & 17.8 & 23.7 & 16.3 & 33.7 & 23.2  \\
        5 &    & BPE-205 & PASM-205 & 21.9 & 24.9 & 17.8 & 24.4 & 16.1 & 33.7 & 23.4  \\
        \cline{3-11}
        6 &    & paraCTC & BPE-205 & 21.3 & 24.7 & 17.3 & 24.0 & 16.2 & 33.1 & 23.1  \\
        \cline{3-11}
        7 &    & paCTC & BPE-205 & 20.5 & 24.1 & 16.5 & 23.3 & 15.3 & 32.6 & 22.4  \\
        8 &    & paCTC  & BPE-3000 & 19.6 & 24.0 & 16.6 & 22.5 & 15.2 & 32.3 & 22.1  \\
        9 &    & paCTC-s & BPE-3000 & 19.9 & 23.6 & 16.4 & \textbf{22.0} & 15.4 & \textbf{31.3} & 21.8  \\
        10 &    & paCTC-us  & BPE-3000 & \textbf{19.8} & \textbf{23.1} & \textbf{15.8} & 22.5 & \textbf{14.8} & 31.4 & \textbf{21.6}\\
		 \bottomrule
	\end{tabular}}
\vspace{-0.4cm}
\end{table*}

Comparing results of system 1-3 in Table \ref{tab:libr}, we see there is no big
difference performance gap between using phonetic-induced PASM and text-induced
BPE as their both CTC/transducer target units. PASM achieves the best result
on the clean test set, while BPE obtains the best one on the other test set.
However, when the proposed PMU modeling methods are applied, both WERs on the
clean and other test sets are greatly reduced. When comparing the results of
ConformerT with conventional BPE-3000 (system 2), even with the basic PMU,
system 6 still achieves relative 8.2\% and 5.0\% WER reductions on the clean and other
sets, respectively. Meanwhile, by comparing system 4 to 6, it's clear that
using PASM as CTC alignments, while larger BPE units as transducer targets is the
best setup for basic PMU, it may due to the fact that, the clear phonetic correspondence
of target units is critical for such time synchronous model.
When comparing system 6 with 7-10, we see continuous WER reduction on the clean test set,
even the performance improvement on the other set is limited. 
Finally, the \texttt{paCTC-us} achieves the best results on the clean testset. 
Compared with the best baseline (system 2), system 10 achieves 12.7\% and 4.3\% 
relative WER reduction on clean and other test set, respectively.

\begin{table}[H]
\vspace{-0.4em}
\renewcommand\arraystretch{1.0}
\caption{WER(\%) on Libri-100hrs clean and other test sets for PMU with paCTC
(Fig.1 (b), Eq.(\ref{eq:pa1})) without optional structure under different setup conditions.
Setup 4 means replacing the PASM-CTC with BPE-194 CTC at $N_{1}$ layers.}
  \label{tab:paCTC}
  \centering
  \scalebox{0.95}{
	\begin{tabular}{l|c|c|c|c|c}
		\toprule
\multirow{2}{*}{ID} & \multicolumn{2}{c|}{\#layer} & \multirow{2}{*}{$\beta$} & \multicolumn{2}{c}{Evaluation} \\
		\cline{2-3} \cline{5-6}
		& $N_{1}$ & $N_{3}$ & & Clean & Other\\
		 \midrule
        1 & 6 &  6  & 0.3  & 9.8 & \textbf{28.0} \\
        2 & 6 &  6  & 0.5  & \textbf{9.7} & 28.4 \\
        3 & 6 &  6  & 0.7  & 10.0 & 28.6 \\
        \hline
        4 & 6 &  6  & 0.5, BPE-194  & 10.0 & 29.4 \\
        \hline
        5 & 3 &  9  & 0.5  & 10.3 & 30.1 \\
        6 & 9 &  3  & 0.5  & 9.8 & 28.6 \\
		 \bottomrule
	\end{tabular}}
    \vspace{-0.3cm}
\end{table}

In fact, before we propose the \texttt{paCTC} with optional structure,
we perform a set of parameter tuning experiments to see how they
affect the \texttt{paCTC} performance. Results are shown in Table \ref{tab:paCTC}.
Setup 1-3, 5-6 are all with the PASM-194 at $N_{1}$ AEncoder block,
and BPE-3000 at the top block. We see that,
$N_{1}=N_{3}=6$ with $\beta=0.5$ achieves relatively stable results.
Furthermore, when we replacing the PASM-194 with BPE-194 for aligning the first $N_{1}$
layers outputs, it obtains worse WERs than setup 2, however, when we compare
it with system 2 in Table \ref{tab:libr}, we still see performance improvements.
This tells us that PASM is more suitable for low-level acoustic
information learning than BPE, and introducing small-to-large target units
progressive representation learning will be helpful. All these observations
lead us to propose the whole structure of \texttt{paCTC} that shown in Fig.\ref{fig:res}(b).

\subsubsection{Results on Accented ASR}
\label{sssec:results}

In Table \ref{tab:accent}, the effectiveness of PMU with its
different variants are examined on the CommonVoice accented ASR task. Different from
the Librispeech task, we find the best vocabulary size of both
PASM and BPE baselines is 205, larger BPE size doesn't result in
better WERs under the ConformerT with single type target units.
And consistent with the findings in Table \ref{tab:libr}, both the in-domain
and out-of-domain performances are continuously reduced by the proposed PMU methods,
such as, system 4 performs better than 5 because PASM is applied on
CTC and while BPE is applied on transducer; \texttt{paraCTC} achieves
better results than basic PMU, and \texttt{paCTC} significantly
outperforms other two PMU variants, especially on the three in-domain test sets.
It's worth noting that, in \texttt{paCTC}, both the intermediate \texttt{BPE-CTC} and \texttt{PASM-CTC}
are with the same 205 vocabulary size. By comparing system 7 with 8, the results also
prove that introducing progressive learning with small to larger target units
is useful. Finally, the \texttt{paCTC} with optional structure achieves the best overall
WERs, compared with the best ConformerT baseline system 2, system 10 produces
relative 7.7\% overall WER reduction on this accented ASR testsets.
Specifically, relative 8.8\%, 7.6\% and 10.2\% WER reductions are
for the in-domain England, Indian and US test sets, 8.2\%, 8.6\% and 7.1\%
WER reductions are for the out-of-domain Australia, Canada and Singapore test set,
respectively.

\section{Conclusion}
\label{sec:conclusion}

In this work, we propose a phonetic-assisted multi-target units (PMU)
modeling approach, to effectively leverage both the phonetic-induced PASM and
conventional text-induced BPE target units modeling for improving
the state-of-the-art Conformer-Transducer end-to-end ASR system.
Three PMU structures are proposed with different implementation of
multi-targets CTC/transducer modeling, including the basic PMU with
PASM and BPE applied to CTC and transducer separately, the PMU with
\texttt{paraCTC} where the PASM and BPE units are also integrated in a parallel way
as CTC's target units, and the PMU with \texttt{paCTC} that uses
BPE units conditioned on the PASM CTC in a progressive representation learning
manner. Results on both the LibriSpeech and accented English ASR taks show that,
the proposed PMU can significantly outperform the conventional BPE-based
Conformer-Transducer E2E ASR system.

\bibliographystyle{IEEEtran}
\bibliography{mybib}

\end{document}